\documentclass[
	letterpaper, % Use US letter paper size
]{jdf}

\author{John O'Connor}
\email{john.oconnor@gatech.edu}
\title{ESCape the ClassRoom:\\A Educational Escape Room in Virtual Reality}

\usepackage{pdfpages}
\usepackage{hyperref}
\begin{document}
%\lsstyle

\maketitle

\begin{abstract}
    Educational Escape Rooms (EER's), through their use of immersive storytelling and practical application of 
    abstract concepts, present a novel new technique for engaging learners in a variety of subjects.  
    However, there is a significant time and materials investment required to build new physical Escape Rooms, and 
    prior attempts to create digital escape rooms have resulted in games that lack the immersive qualities that 
    make physical escape rooms so compelling.  
    This paper presents ESCape the Classroom, a web framework for creating virtual reality educational escape 
    rooms (VR EERs) that can be delivered to any web-connected device.
    The framework is equipped with essential tools to design and deploy intricate, multi-room VR escape 
    experiences using HTML and WebComponents. It is designed to be used by educators with rudimentary programming skills, 
    eliminating the need for advanced game programming or development expertise. VR EERs created with this platform can 
    be published online as WebXR sites that are compatible with a broad spectrum of VR hardware, including the Meta Quest 3, 
    allowing educators to share the experiences they create while bypassing the need for additional software installations 
    on devices. This paper will present the
    design and implementation of ESCape the Classroom, and discuss the potential for this platform to be used in
    educational settings.

\end{abstract}
\section{Introduction}

Escape rooms, known for their enigmatic puzzles and thrilling time constraints, have soared in popularity as a form of entertainment. 
When designed with educational purposes in mind, these escape rooms can align game elements with specific 
learning objectives, creating an engaging and dynamic educational tool. Such educational escape rooms have 
been shown to be effective learning platforms, according to research by 
\cite{gordon_escape_2019} and \cite{gordillo_evaluating_2020}. However, the physical versions of these rooms 
are often expensive to create and maintain \citep{fotaris_room2educ8_2022}. Digital versions, on the other hand, have generally lacked physical 
immersion due to their reliance on online forms \citep{anton-solanas_evaluation_2022}. Furthermore, both physical and 
digital escape rooms require significant time and effort to design, particularly for educators without 
game design expertise \citep{fotaris_designing_2023}.

Virtual reality (VR) offers a promising alternative. Games such as "Among Us VR" and "Keep Talking and Nobody 
Explodes" illustrate the potential for engaging puzzle games within VR environments, benefiting from VR's unique 
features like player-controlled cameras \citep{nyyssonen_exploring_2021}. However, these games typically involve 
simple, non-educational puzzles and do not support the creation of customizable learning scenarios due to 
their closed-source nature.

This gap underscores the need for a VR-based educational escape room platform that replicates the immersive 
experience of physical rooms and supports the development of educational content without the associated high 
costs. Advances in immersive Mixed Reality (XR) technologies and the availability of affordable XR headsets, 
such as the Meta Quest 3, are making such immersive experiences increasingly feasible and accessible in 
educational settings. Additionally, the integration of Large Language Models in the design and evaluation 
phases of educational escape rooms has been explored, showing potential for these models to enhance formative
 assessments, provide personalized feedback to learners \citep{pinto_large_2023}, and speed up the design process \citep{nye_generative_2023}.

\section{The Need for Virtual Reality in Educational Escape Rooms}

Educational Escape Rooms (EERs) have proven to be a dynamic tool for engaging learners by incorporating 
immersive storytelling and the practical application of abstract concepts \citep{lopez-pernas_escapp_2021}. 
Traditional physical escape rooms excel in providing this immersive experience, creating a sense of presence 
that is both engaging and memorable\citep{lopez-pernas_examining_2019}. However, the construction of 
these physical spaces demands considerable resources, including time, money, and physical 
materials\citep{fotaris_room2educ8_2022}, which can be prohibitive. Additionally, while digital escape 
rooms have been developed as an alternative \citep{anton-solanas_evaluation_2022}, many of these offerings are delivered 
as forms or quizzes, which lacks the depth of immersion that makes physical escape rooms so compelling \citep{quek_educational_2024}.

This gap in digital escape room design—where the physical sensations and real-time interactions that make 
traditional escape rooms profoundly engaging are absent—highlights a significant opportunity within educational 
technology. Virtual Reality (VR) emerges as a compelling solution to this problem. VR uniquely allows the 
creation of spatially navigable environments that users can interact with in a manner that closely mirrors 
real-world interactions \citep{chukusol_virtual_2022}. This technology can simulate the immersive aspects of physical rooms, including the 
ability to explore, manipulate objects, and solve puzzles in a seemingly tangible world, all without the 
associated logistical constraints and overheads.

Furthermore, VR technology has reached a level of maturity and accessibility that makes it a practical option for
 educational settings \citep{wohlgenannt_virtual_2020}. The latest VR hardware, such as the Meta Quest 3, offers high-quality immersive 
 experiences at a price point that is becoming increasingly feasible for educational institutions. 
 These devices are also becoming easier to use, with intuitive interfaces that both educators and students 
 can navigate without prior experience in virtual environments \citep{kao_evaluating_2021}.

The potential of VR to bridge the engagement gap in digital educational escape rooms is partially supported by research \citep{shute_stealth_nodate}
indicating that immersive VR environments can potentially enhance learning outcomes \citep{gordon_escape_2019}. Studies have 
shown that VR can improve retention rates in some cases \citep{smith_learning_2016}. If educators
can create compelling, immersive educational experiences, they can capture the benefits of physical escape 
rooms but in far more scalable and cost-conscious way. These escape rooms can be iteratively adapted to meet 
diverse educational needs without substantial ongoing costs.

% The full breakdown is included in this document as \href{Appendix A}{Appendix A}.
\section{ESCape the ClassRoom Framework Overview}

The "ESCape the Classroom" framework integrates several technologies and custom components to create immersive and interactive virtual reality educational escape rooms (VR EERs). 

\subsection{Game State Management}

This framework uses X-State JS and and a custom game-state AFrame component to manage game states effectively. 
It dynamically adds rooms and puzzles to the state machine by defining game states as entities and attaching the game-state 
component to those entities.  This is done by adding the <a-entity> tag for the puzzle in the HTML file and adding the component as an attribute to that entity.

\begin{verbatim}    
    <a-entity id="puzzle1" 
    game-state="type:puzzle; name:puzzle1; room:room1"></a-entity>
\end{verbatim}

Other components can listen for game-state-event events from the VR scene to facilitate transitions based on 
player interactions.

The state management system is robust, emitting a game-state-updated event whenever the state changes, 
which allows other components within the VR environment to react dynamically to new states. 

\begin{verbatim}
    <a-gltf-model id="stage" 
    ...
    hide-in-state="state: running.debriefing.debriefingPlay; 
    showOtherwise: true"
    ...
  ></a-gltf-model>

\end{verbatim}
All game states 
are stored directly on the "scene" object as a component, making them easily accessible for updates and queries
 throughout the game.  Since A-Frame entities use HTML, they can be selected using the standard DOM API's like \texttt{document.queryselector}.

 \begin{verbatim}
    document.querySelector("a-scene").emit("game-state-event", "loaded");
    document.querySelector("a-scene").addEventListener(
        "game-state-event", (event) => {
        // Handle game state changes
    });
    document.querySelector("a-scene").addEventListener(
        "game-state-updated", (event) => {
        // Handle game state updates
    };

\end{verbatim}

\subsection{Navigation and Movement}

Navigation within the virtual space is managed through a navmesh generated using the recast navmesh library. 
This mesh, created from GLTF models, defines where players can move, enhancing the realism of the VR experience. 

\begin{jdffigure}
    \includegraphics[width=6in]{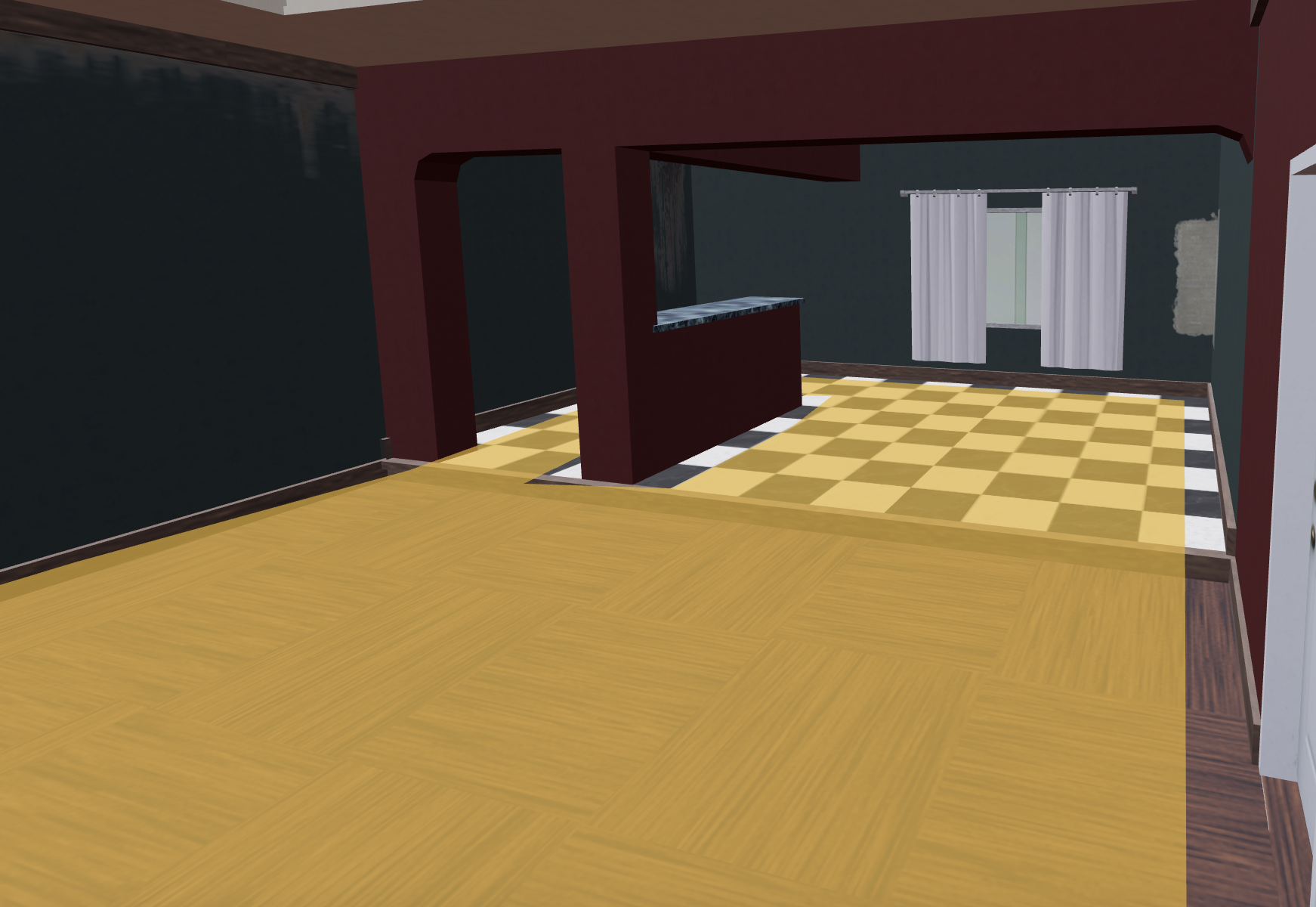} \\
    \captionof{figure}{The apartment model from the included demo, showing the visible navmesh}\label{fig:navmesh}%
    \end{jdffigure}

Movement is restricted to designated areas via the simple-navmesh-constraints AFrame component. The navmesh adapts 
to gameplay, creating or removing barriers in response to state changes, such as unblocking parts previously 
inaccessible or altering the visibility of room elements like doors using the gltf-hide AFrame component.

\begin{verbatim}
    <a-entityid="cameraRig"
    simple-navmesh-constraint="navmesh:.navmesh;exclude:.navmesh-hole;"
    ...
  >

  <a-gltf-model id="stage" 
    src="#apartment" 
    ...
  gltf-hide="parts:apartmentDoor001,apartmentDoor002,apartmentDoor"
></a-gltf-model>

\end{verbatim}

The ESC-CR Custom Components enrich the framework's capabilities; these include the 
Base Component which wraps the Web2VR lifecycle into a web component and the 
\texttt{esc-html-panel} that converts HTML content into interactive AFrame components.

\begin{verbatim}
    <esc-html-panel id="panel" 
    ...
    <h1>ESCape the ClassRoom Demo</h1>
    <a class="btn btn-primary" href="/apartment.html">
        Enter Geometry Game
    </a>
    ></esc-html-panel>

\end{verbatim}

The image generated by this code snippet is below:

\begin{jdffigure}
    \includegraphics[height=3in]{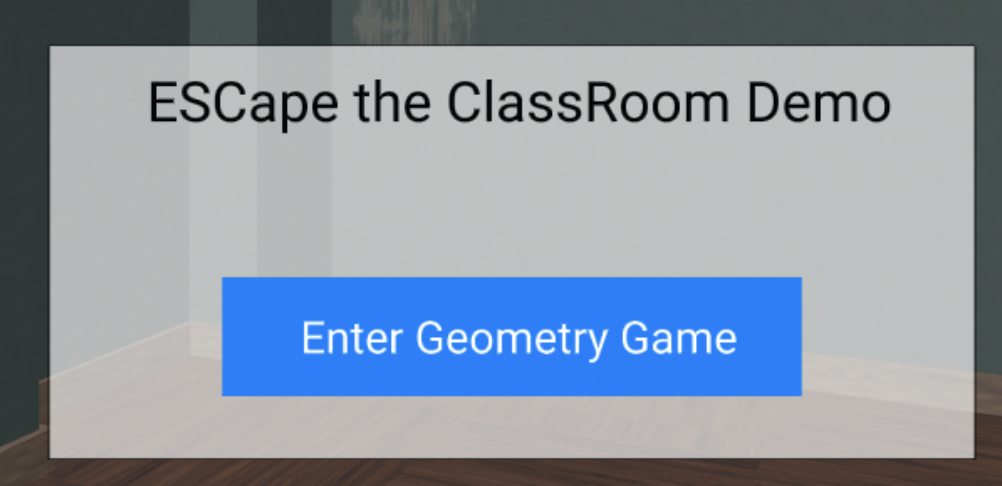} \\
    \captionof{figure}{An example of an HTML panel rendered into 3D}\label{fig:htmlpanel}%
    \end{jdffigure}

The watch component combines multiple techniques, rendering an \texttt{esc-html-panel} 
component onto a gltf model to create a virtual wristwatch that counts down time, 
adding a layer of urgency to the escape room challenges. 

\begin{jdffigure}
    \includegraphics[height=3in]{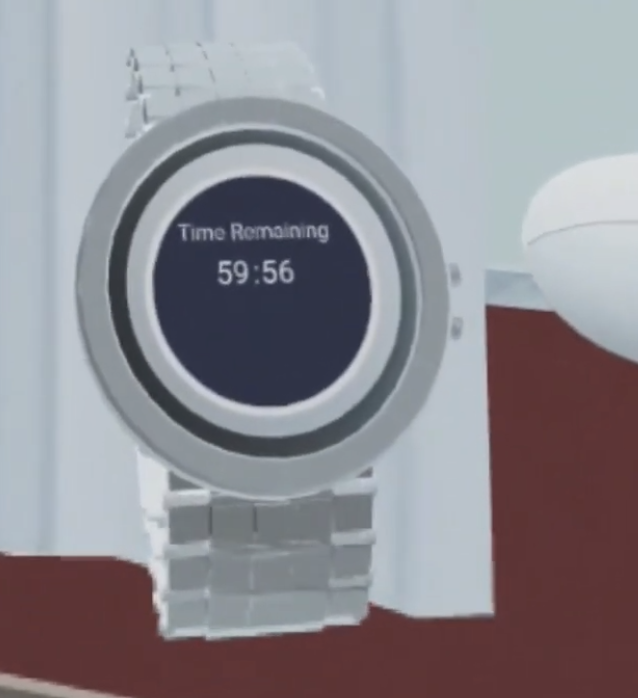} \\
    \captionof{figure}{The countdown wristwatch that combines multiple techniques in 2D and 3D space}\label{fig:htmlpanel}%
    \end{jdffigure}

This is done by creating the HTML panel and attaching it to an entity that 
uses the watch model.

\begin{verbatim}
    <esc-watch id="mainWatch" 
        components='{"game-clock": true...}
        settings='{"parentSelector": "#watchEntity"}'>
        <div class="watch">
            <div class="time_header">Time Remaining</div>
            <div class="time">
                <span class="minutes">60</span>:
                <span class="seconds">00</span></div>
            </div>
        </div>
    </esc-watch>
    ...
    <a-entity laser-controls="hand: left"...>
        <a-entity id="watchEntity" gltf-model="#watchModel"...></a-entity>
    </a-entity>
\end{verbatim}

\section{Design and Implementation of VR EERs}

The "ESCape the Classroom" framework provides educators with an intuitive and robust platform 
for designing and implementing multi-room, multi-puzzle virtual 
reality (VR) educational escape rooms. This section delves into the 
technical details of the framework, highlighting the design of the 
user interface, interaction mechanisms, and the integration of 
educational content within the VR environment.

\subsection{Implementing an EER with ESCape the Classroom}

Creating an EER with ESCape the Classroom starts by creating an HTML page and importing the application libraries.

Once the libraries are imported, the VR scene is defined using standard HTML tags, with the A-Frame library providing the necessary components for creating the 3D environment. The scene is populated with rooms, puzzles, and interactive objects, each defined as entities with specific attributes and behaviors.

Teachers can download 3D models from many sources, such as sketchfab, and import them into the scene using the GTLF standard format.

ESCape the classroom provides multiple A-Frame components that accelerate
creation of an escape room, such as the game-state component, which manages the state of the game, 
and the simple-navmesh-constraints component, which restricts player 
movement to specific areas.  The framework also includes an open source navmesh
generator based on \texttt{recast} making it easy to generate navmeshes.

\subsection{Interaction Design}

Interaction design within the framework is centered around providing a 
realistic and engaging user experience that promotes learning. 
The VR escape rooms support a variety of interaction types, 
from simple point-and-click to more complex manipulation tasks, 
such as assembling or repairing objects to solve puzzles. These 
interactions are facilitated through the use of VR controllers, 
which allow users to perform gestures like grabbing, pulling, 
turning, and pressing within the virtual environment.

To enhance the educational value of the escape rooms, interactions are 
often linked to educational outcomes. For example, a puzzle might 
require learners to solve a math problem to find the combination to a 
lock or to perform a sequence of chemical reactions to safely dispose 
of a hazardous material. These tasks encourage critical thinking and 
problem-solving skills, aligning the VR activities with curriculum 
objectives.

ESCape the ClassRoom provides components to help facilitate puzzle creation and 
interaction within the room, and it's use of the WebComponent standard means a community
of educators using this framework could share components with each other.

Since the framework is built on top of A-Frame and the standard web Document-Object Model, 
existing knowledge of web development can be leveraged to create more complex interactions.

\subsection{Integration of Educational Content}

Educational content is integrated into the VR escape rooms through 
interactive puzzles and narrative elements that align with learning 
objectives. The framework supports the embedding of rich media content, 
including text, images, audio, and video through the use of standard HTML tags.
The framework converts these into 3D objects, and maps the VR events to 
the DOM events, allowing for seamless integration of multimedia content.
These can be used to provide 
instructions, background information, and feedback. This multimedia 
approach caters to different learning styles and helps to keep students 
engaged and motivated.

Teachers can leverage existing web metrics and analytics API's by attaching
them to the DOM events generated by ESC-CR, allowing educators to track progress
 and evaluate learning outcomes using existing tools and APIs. 
 These tools can collect data on how students interact with different 
 puzzles and how they solve problems, providing insights into their 
 understanding and areas where they may need further guidance or 
 support.
 
\section{Publishing and Compatibility of VR Educational Escape Rooms}

The "ESCape the Classroom" framework not only facilitates the creation of 
virtual reality educational escape rooms but also addresses the 
challenges associated with publishing and compatibility across 
various platforms. This section explores the mechanisms through 
which VR escape rooms are published and how they ensure compatibility 
with a broad spectrum of VR hardware, including low-cost solutions 
like the Meta Quest 3.

\subsection{Publishing VR Escape Rooms}

Once an escape room is designed and implemented using the "ESCape the Classroom" 
framework, the process of making it accessible is straightforward and 
efficient. Educators can publish their VR escape rooms directly to the 
web as WebXR HTML sites. This method of publishing is significant 
because it eliminates the need for users to download and install 
specific applications or software. Instead, users can access the VR 
experiences directly through their web browsers, which dramatically 
simplifies distribution and access.

The framework utilizes standard web technologies for deployment, which 
ensures that updates and modifications to the VR escape rooms can be 
rolled out quickly and seamlessly. This is particularly beneficial 
for educational settings where content may need to be refined based on 
feedback or changing educational needs.

\subsection{Ensuring Compatibility Across Devices}

Compatibility is a cornerstone of the "ESCape the Classroom" framework. 
The VR escape rooms created with this platform are designed to be 
compatible with a wide range of VR hardware. This inclusivity is 
achieved through the use of responsive design principles and adaptive 
scaling technologies that adjust the VR content according to the 
capabilities and specifications of the hardware being used.

The framework uses WebXR, a specification published and maintained by the World Wide Web Consortium \citep{maclntyre_thoughts_2018}.
Any web browser that supports this specification can potentially be used
to view and interact with ESC-CR experiences.  The platform has specifically 
tested and known working on the Meta Quest 2 and 3, which are known for its affordability and ease of 
use in educational environments. By ensuring that the VR escape 
rooms work seamlessly on devices like these, the framework broadens the 
potential user base, making these educational tools accessible to a 
wider audience of learners.

\section{Discussion and Future Work}

The introduction of the "ESCape the Classroom" framework brings 
transformative potential to educational settings through virtual 
reality (VR). This section will delve into the broader implications 
for VR in education, address current issues, and propose directions 
for future work to enhance the platform's impact and reach.

\subsection{Future Work}
\subsubsection{Multi-player Games}

Future enhancements to the framework could focus on improving the 
capabilities for multi-player experiences, allowing groups of
students to collaborate or compete within the same VR space. 
This would encourage teamwork and communication, fostering 
social learning and problem-solving skills.

\subsubsection{Teacher Observers}

Integrating features that allow teachers to observe and interact with 
students in real-time within the VR environment could greatly enhance 
the educational experience. This would enable instructors to provide 
immediate guidance and feedback, tailor interventions as needed, and 
better understand the dynamics of student interactions.

\subsubsection{Formative Assessments and Dynamic Puzzles}

Developing more sophisticated formative assessments and dynamic puzzles 
that adapt to a student's performance could personalize learning and 
increase engagement. These tools could automatically adjust the 
difficulty of tasks based on the student's progress, providing a 
customized learning path for each user.

\subsubsection{Evaluation as a Teaching Tool}

Further research could be conducted on the effectiveness of VR as a 
teaching tool, particularly in terms of its ability to improve learning 
outcomes compared to traditional methods. This could include studies 
on retention, comprehension, and the ability to apply learned skills 
in practical settings.

\subsection{Issues}
\subsubsection{Disability Accessibility}

While VR offers novel educational opportunities, significant work is 
needed to make these technologies accessible to students with 
disabilities. This includes the development of VR content and 
controls that are adaptable for various physical and sensory 
impairments, ensuring that all students can benefit from immersive 
educational experiences.

\subsubsection{Access to VR Headsets}

The cost and availability of VR headsets remain a significant barrier 
to widespread adoption in educational settings. Strategies to address 
this issue could include partnerships with VR hardware manufacturers, 
grants, and funding initiatives aimed at increasing access to necessary 
technology in schools, especially in under-resourced areas.

\section{Conclusion}

The "ESCape the Classroom" framework represents a significant leap 
forward in the integration of virtual reality (VR) technologies within 
educational settings. By providing an innovative platform for the 
creation and dissemination of educational escape rooms, this framework 
offers a compelling blend of engagement, interactivity, and immersive 
learning experiences that traditional educational methods often lack.

Throughout this paper, I have explored the technical capabilities of 
the framework, including the methods for creating experiences using HTML,
the tools it offers to enable dynamic 
management of game states, and the seamless publishing process that 
ensures broad compatibility across various VR devices. The framework's 
emphasis on accessibility and its ability to deliver educational 
content directly via WebXR sites make it a powerful tool in the 
modern educator's arsenal, breaking down some traditional 
barriers to technology adoption in education.

However, as with any emerging technology, there are challenges to 
overcome. Issues such as disability accessibility and the affordability 
and availability of VR headsets pose significant hurdles to universal 
adoption. Addressing these concerns requires ongoing effort and 
innovation, as well as partnerships between educational institutions, 
technology providers, and policymakers.

Looking ahead, the potential for future enhancements such as 
multi-player functionalities, real-time teacher observation tools, 
and adaptive learning puzzles promises to further enhance the 
educational value of VR escape rooms. Continued research into the 
effectiveness of VR as a teaching tool will also be critical in 
validating and refining these technologies, ensuring they meet the 
diverse needs of learners and educators alike.

\section{References}

\printbibliography[heading=none]

\appendix

\end{document}